# Different responses of the upper ocean to typhoon Namtheun

Liang SUN[1,2], Yuan-Jian Yang[1], Yun-Fei Fu[1,2]

1. Laboratory of Satellite Remote Sensing and Climate Environment, School of Earth and Space Sciences, University of Science and Technology of China, Hefei, Anhui, 230026, China;

2. LASG, Institute of Atmospheric Physics, Chinese Academy of Sciences, Beijing 100029, China

Corresponding author address:

Liang SUN

School of Earth and Space Sciences, University of Science and Technology of China

Hefei, Anhui 230026, China

Phone: 86-551-3600175; Fax: 86-551-3606459

Email: sunl@ustc.edu.cn; sunl@ustc.edu

**Abstract:**

The responses of the upper ocean to typhoon Namtheun in July 2004 are investigated by sea surface measurements and vertical profiles. Pre-typhoon ocean environment played an important role in this case. There were two extreme cooling regions located at cyclonic eddy A and typhoon wake B. Although the magnitudes of SST cooling at A and B were similar, other physical and biophysical responses were quite different. Combining multi-satellite data with vertical profile data, it is found that the upwelling dominated the responses at A and the vertical mixing dominated the responses at B. This study implies that to insight into the ocean surface responses to typhoon, the subsurface dynamics need to be analyzed via both the in situ and satellite-based observations, and the physical and biological models.

.

## 1. Introduction

Typhoon (or hurricane, tropical cyclone), when passing over the ocean, has both physical [Price 1981; Stramma et al. 1986; Price 1994; Emanuel, 1999; Chen et al., 2003; Liu et al., 2008] and biophysical [Behrenfeld et al., 1997; McGillicuddy et al., 1999; Subrahmanyam et al., 2002; Lin et al., 2003; Babin et al., 2004] impacts on the ocean. As a result, vertical mixing, entrainment and upwelling injured colder subsurface water [Price, 1981; D'Asaro, 2002; Walker et al., 2005], made cold wake at sea surface, and leaded to phytoplankton blooming along typhoon track, especially in oligotrophic waters [Babin et al., 2004].

The mechanisms of blooming were commonly considered to be related to upwelling of nutrition [Subrahmanyam et al., 2002; Lin et al., 2003; Tang et al., 2004a, 2004b; Shi and Wang, 2007], cold-core cyclonic eddies [Biggs and Muller-Karger, 1994; Walker et al., 2005; Gierach and Subrahmanyam, 2008], entrainment in mixed layer [Davis and Yan, 2004; Zheng and Tang, 2007], and vertically mixed chromophoric dissolved organic matter (CDOM) from deeper water [Hoge and Lyon, 2002; Shang et al., 2008]. Since the phytoplankton blooming, or chlorophyll-a (chl-a) enhancement, consequently appeared accompanied with sea surface cooling after typhoon, the inverse correlation ship between SST cooling and chl-a enhancement has been widely documented [Subrahmanyam et al., 2002; Lin et al., 2003; Babin et al., 2004; Walker et al., 2005], and the SST cooling was even taken as a proxy of phytoplankton blooming [Gierach and Subrahmanyam, 2008]. However, the classical concept [Price 1981] about ocean response to typhoon was proven to be invalid in some cases, where there were some pre-existing eddies [Gierach and Subrahmanyam, 2008]. And the inverse relationship between SST cooling and chl-a enhancement was invalid in offshore wake of typhoon Hai-Tang [Chang et al., 2008].

Besides, Babin et al. (2004) guessed that entrainment of phytoplankton from the deep chlorophyll maximum might lead to chl-a enhancement, which was plausible according to some sea surface observations [Davis and Yan, 2004; Walker et al., 2005; Gierach and Subrahmanyam, 2008]. However, such guess has not been clearly concluded, due to the lack of interdisciplinary subsurface data sets [Babin et al., 2004]. Both the invalid concept about ocean response and the uncertain guess about chl-a enhancements imply that further investigation on ocean response is required. To reveal the mechanism of chl-a enhancement, the vertical distribution of nutrients and phytoplankton are the primary factors [Babin et al., 2004].

In this study, the responses to Namtheun are presented by using multi-satellite data, in situ cruise data and Argo floats profiles. With the vertical profiles, the SST cooling and chl-a enhancement due to entrainment and upwelling are investigated. It is found that pre-typhoon ocean environment played an important role in the ocean responses to typhoon.

## 2. Observation Data and Background

### 2.1 observation data

Typhoon track data, taken by every 6h, including center location, central pressure, and maximum sustained wind speeds (MSW), were obtained from Shanghai Typhoon Institute (STI) of China Meteorological Administration (CMA). With spatial resolution of 0.25°×0.25°, sea surface wind vectors and sea surface temperature data, which were respectively derived from QuikSCAT (Quick Scatterometer) and the tropical rainfall measuring mission (TRMM) microwave image (TMI), are produced by the Remote Sensing Systems (www.remss.com). These data were used to calculate upwelling speed and thermocline displacement [Price et al., 1994]. The sea surface height anomaly (SSHA) data were produced and distributed by AVSIO (Archiving, Validation and Interpretation of Satellite Oceanographic data). Merged Level 3 daily chl-a data, with a spatial resolution of 9 km from two ocean color sensors (MODIS and SeaWiFS), were produced and distributed by the NASA Goddard Space Flight Center's Ocean Data Processing System (ODPS). Vertical profile data for chl-a and nutrient were from Japan Oceanographic Data Center (JODC). And the Argo float profiles were extracted from the real-time quality controlled Argo data base of China Argo Real-time Data center.

*2.2 backgrounds before Namtheun*

Typhoon Namtheun was a classical typhoon formed over the subtropical Pacific Ocean on 07/24, 2004. Latter, it intensified to a super category 4 typhoon (maximum wind: 115 knot) with rapidly moving speed (~6 m/s) on 07/26. Then, Namtheun lingered westward at a near stationary slow speed (~2 m/s) between 07/28 and 07/30 (Fig. 1). After that, Namtheun moved quickly (~6 m/s) towards northwest, and it finally landed Shikoku island and weakened gradually. Meanwhile, both the geostrophic current and the sea surface height anomaly prior to Namtheun are shown between 07/17 and 07/23, where there was a pre-existed cyclonic eddy marked A, which was warm-core from the vertical temperature profile at station 023 (Fig 1b). And the pre-Namtheun Argo floats are also marked with black triangles.

The vertical profiles prior to Namtheun, measured at cruise stations (Fig. 1a, Table 1), are shown in Fig. 1b. The chl-a intensity increases from the sea surface with depth till it approaches its maximum at critical depth, then decreases with depth in the deeper ocean. The chl-a maximum depth for station 13, 17, 23 and 46 was 125 m, 100 m, 75 m and 50 m, respectively, below which there was insufficient sunlight, above which there was insufficient nutrition. However, the nitrate and the phosphate only exist below a certain depth (chl-a maximum depth), below which the nutricline extends to the deep ocean. The deeper the water is, the larger the nutrition intensity is. Thus the constraint for chl-a intensity is nutrition and sunlight above and below the chl-a maximum depth, respectively.

Moreover, the measurements by SeaWiFS and in situ observations are compared (Table 1). The in situ chl-a, which was the vertical average of vertical integration of chl-a intensity within whole euphotic zone (depth ~200 m), was about 10% larger than that by the SeaWiFS data, so they agreed well with each other [O'Reilly et al., 1998, 2000]. As chl-a concentration by SeaWiFS is not the surface chl-a intensity but the vertical average chl-a intensity, entrainment of phytoplankton from the deep chl-a maximum, though taking chl-a maximum shallower and changing chl-a profile, can't change the vertical average chl-a alone without entrainment of nutrient, thus the former guess [Babin et al., 2004] seems invalid in this case.

## 3. Results

### *3.1 surface responses to Namtheun*

Typhoon Namtheun leaded sea-level decrease, sea surface cooling and chl-a enhancement in the ocean. The physical responses, including sea level decrease (Fig. 2a) and sea surface cooling (Fig. 2b), are presented. Prior to Namtheun, there were dominated by the warm water with temperature T>27°C (Fig. 2b). After Namtheun, there were two extreme cooling regions (~5°C of SST decrease), which located at the cyclonic A and typhoon wake B (Fig.2b), respectively. Although the magnitudes of SST cooling at A and B were similar, the SSH changes were something different. In Fig 2a, there was widely upwelling, and the SSH decreased during the passage of Namtheun. However, the SSH decreased in region A (with maximum of ~50 cm) was much deeper than that in B (with maximum of ~10 cm). According to Price [1994], the thermocline upwelled about 65 m and 12 m at A and B, respectively, which agreed well with former investigations [Walker et al., 2005]. Although the SST cooling at A and B seemed similar, the consequences of cooling were different from the biological responses.

The biological response, indicated by chl-a concentration, is depicted in Fig. 2c. Prior to Namtheun, there was a typical summer condition of chl-a concentration, predominantly < 0.1 mg/m$^3$ in the offshore region, and >0.1 mg/m$^3$ in the nearshore region, especially in coastal waters. The chl-a concentration enhanced after typhoon's passage. In the cyclonic eddy A, the chl-a concentration mean increased dramatically from 0.14 mg/m$^3$ (pre-Namtheun) to 0.26 mg/m$^3$ (post-Namtheun). Contrast that, the chl-a concentration mean increased slightly from 0.088 mg/m$^3$ (07/24-26) to 0.104 mg/m$^3$ (08/01-02), then damped to 0.075 mg/m$^3$ (08/08-10) later in the wake B.

The responses at sea surface, indicated by SST and chl-a at A and B, are shown in Fig. 2d. At region A, the SST cooling appeared gradually and lasted for about a week, and the phytoplankton blooming was inversely correlated to sea surface cooling, however, with 3-4 days lag (Fig 2d). This lag time is consistence with former investigations [Babin et al., 2004; Walker et al., 2005; Zheng and Tang, 2007]. Comparing with these changes at region A, the SST cooled very rapidly but returned back immediately, and chl-a concentration seldom changed at region B, which were something like what had happened after typhoon Hai-Tang [Chang et al., 2008]. The different physical and biophysical responses between A and B can be understood further from the subsurface process.

*3.2 subsurface responses to Namtheun*

The subsurface responses to Namtheun, including entrainment and upwelling, can be directly seen from the snapshots of Argo float profiles (Fig. 3). There were two Argo floats right near the Namtheun track (Fig. 1a), where the mixing depth was significantly shallower than euphotic zone depth, while the upwelling occurred obviously below euphotic zone (Fig 3a, b, c). Both floats recorded dramatically cooling after Namtheun (Table 2), which can also be seen from the both temperature profiles (Fig 3b, c) and the SST time series (Fig. 3d). Both the vertical mixing and upwelling at float 2900139 were significantly larger than those at float 2900141 (Fig 3b, c; Table 2). As a result, the SST cooling was weaker and lasted longer at float 2900139 than that at float 2900141 (Fig. 3d), even though float 2900141 was much closer to typhoon track. Moreover, the chl-a enhancement at float 2900139 was relatively larger than that at float 2900141 (Fig. 3d). Noting that SSHA and temperature profiles showed both weaker mixing and weaker upwelling at float 2900141, this might also be the reason why the chl-a enhancement at float 2900141 was weaker than that at float 2900139.

## 4. Discussion

As mentioned above, the physical and biophysical responses were different at region A and region B. What is the reason?

At first, we exclude the translation speed of the typhoon and the entrainment of phytoplankton from the deep chl-a maximum. It is obvious that entrainment couldn't change the vertical average chl-a alone, especially when the mixing zone depth was significantly shallower than euphotic zone depth (Fig 3), which might also be the stories happened at wakes of typhoon Hai-Tang [Chang et al., 2008] and Wilma [Gierach and Subrahmanyam, 2008]. And the impact of different translation speeds on SST was very small, as both moving speeds (3.93m/s vs. 3.73m/s) and maximum wind speeds (40 m/s vs. 37 m/s) of Namtheun were similar over region A and B. Noting that the upwelling was quite stronger at A (Fig. 2a), it seems that the pre-existed cyclonic eddy at region A played an important role in this process, which agreed well with former investigations [Walker et al., 2005; Gierach and Subrahmanyam, 2008; Shang et al., 2008].

Then by combining the multi-satellite data and vertical profiles, it is concluded that the upwelling dominated the responses at region A and the vertical mixing dominated the responses at region B. For the region A, the SSH and SST decreased dramatically, and the chl-a enhanced significantly after the typhoon due to upwelling. However, the mixing played minor role in this process. From the temperature profile of station 023 (at region A), the subsurface water was much warm (Fig 1b, average 26.34°C over 100 m depth), so the vertical mixing alone can't provided such strong cooling occurred at region A. And the time series of SST and chl-a (Fig. 2d) also indicated persistent upwelling at this region. Contrast that, the upwelling was very weak from SSH data, so vertical mixing dominated the changes at region B, though both upwelling and mixing were weak here. This point is also supported by profile of float 2900141 (at region B), the subsurface water was colder at float 2900141 (Fig 3a, average 19.92°C over 100 m depth). The shallower mixing only cooled sea surface but seldom changed chl-a concentration at region B, which is consistent with the time series of SST and chl-a at region B (Fig 2d), and is also consistent with the above conclusion, i.e., entrainment can't change the vertical average chl-a alone without entrainment of nutrient.

In a word, the upwelling was dominated at region A and the vertical mixing was dominated at region B, which is the reason why the responses were different at the places. In this case, the pre-typhoon environments , especially subsurface conditions, played essential role in ocean responses. It implies that the subsurface environment also played important role in air-sea interactions [Chen et al., 2003].

## 5. Conclusion

In summary, the responses of the upper ocean to typhoon Namtheun were investigated by observations of SSW, SST, chl-a, nutrient and Argo floats, etc. There were two extreme cooling regions located at cyclonic eddy A and typhoon wake B. Although the magnitudes of SST cooling at A and B were similar, other responses were something different. By combining the multi-satellite data with vertical profiles, it is found that the upwelling dominated the responses at region A and the vertical mixing dominated the responses at region B. Thus the pre-typhoon ocean environment played an important role in the ocean response to typhoon. This study implies that to insight into the ocean surface responses to typhoon, the subsurface dynamics need to be analyzed via both the in situ and satellite-based observations, and the physical and biological models.

**Acknowledgements:**

We thank STI for providing typhoon track data, NASA's Ocean Color Working Group for providing merge MODIS and SeaWiFS data, Remote Sensing Systems for TMI sea-surface temperature and QuikScat wind-vector data, AVISO for SSHA data and geostrophic current data, JODC for Ocean current data and Physical-chemical oceanographic data, and China Argo Real-time Data center for float profiles.

This work is supported by the National Basic Research Program of China (No. 2007CB816004), the National Foundation of Natural Science (Nos. 40705027, 40730950 and 40675027).

Table 1 In situ observations, where IC, AC, SC, NA represent "vertical integration of chil-a intensity", "average of vertical integration of chil-a intensity", "chl-a concentration by SeaWiFS", and "not available", respectively.

| Stations (Location) | Date | Depth (m) | IC(mg/m$^2$) | AC (mg/m$^3$) | SC(mg/m$^3$) |
|---|---|---|---|---|---|
| 013 (137°E,30°N) | 07/16 | 201 | 28.32 | 0.141 | 0.128 |
| 017 (137°E,31°N) | 07/16 | 202 | 25.64 | 0.127 | NA |
| 023 (137°E,32°N) | 07/17 | 202 | 27.63 | 0.137 | NA |
| 046 (135°E,29°N) | 07/27 | 305 | 33.45 | 0.109 | 0.098 |

Talbe 2. Observations by Argo Floats, where TD, MLD, MLT and MLS represent thermociline displacement, mixed layer depth, mixed layer temperature, mixed layer salinity, respectively.

| Argo floats | Date (Position) | TD (m) | MLD (m) | MLT(℃) | MLS (psu) |
|---|---|---|---|---|---|
| 2900139 | 07/27 (141.1°E,32.5°N) | | 25 | 27.60 | 34.828 |
| | 08/01 (140.9°E,32.7°N) | 70-80 | 50 | 24.17 | 34.707 |
| 2900141 | 07/23 (142.7°E,30.5°N) | | 10 | 27.594 | 34.749 |
| | 07/28 (142.6°E,30.4°N) | 40-50 | 25 | 26.718 | 34.825 |
| | 08/02 (142.7°E,30.6°N) | 20-30 | 30 | 23.893 | 34.716 |

**Figure 1. (a)** Study area (long short dashed Box) and track of Typhoon Namtheun. Typhoon center positions every 6 h are indicated by open circles (circle diameter: maximum sustained wind speeds [MSW, m/s], central pressure [hPa]). Pre-Namtheun geostrophic current and SSH are shown between 17 and 23 Jul. The black triangles and black circles indicate the pre-Namtheun position of Agro floats and cruise stations, respectively. **(b)** The in situ vertical profiles of Chl-a, nitrate, phosphate, and temperature at cruise stations.

**Figure 2.** The images of SSH (a), SST (b), and chl-a (c) before, during and after Namtheun, where open circles represent typhoon track and the notation "07/21~27" represents the average value from 07/21 to 07/27, etc., and the time series of Chl-a concentration and SST at both A and B (d).

**Figure 3.** Vertical profiles of temperature (T: squires) and salinity (S: curves) measured by (a) 2900141 on 07/23 (dashed) and 07/28 (solid), (b) 2900141 on 07/23(dashed) and 08/02 (solid), and (c) 2900139 on 07/27 (dashed) and 08/01 (solid), respectively, and (d) the time series of Chl-a and SST at the floats 2900139 and 2900141.

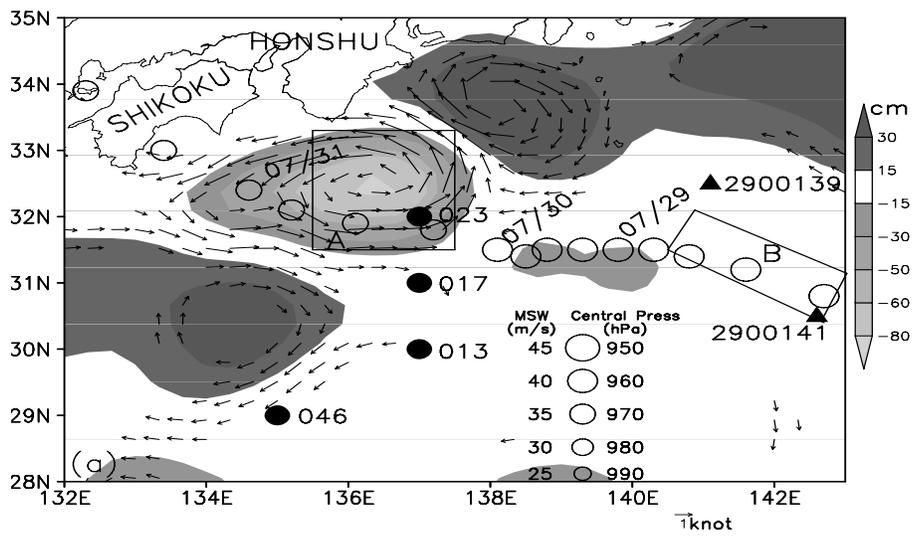

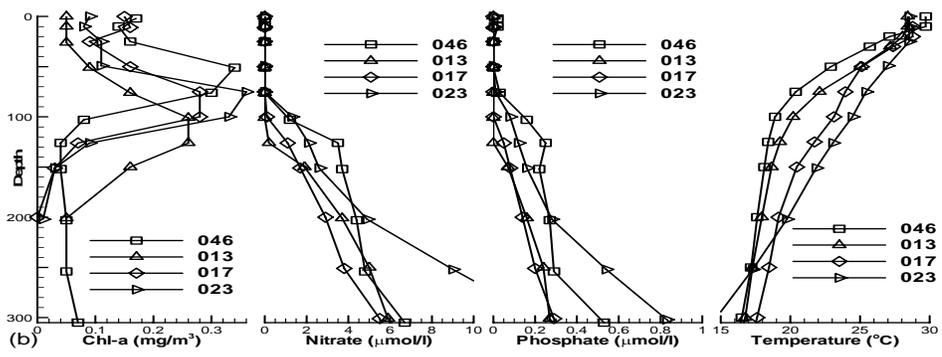

3 Figure 1

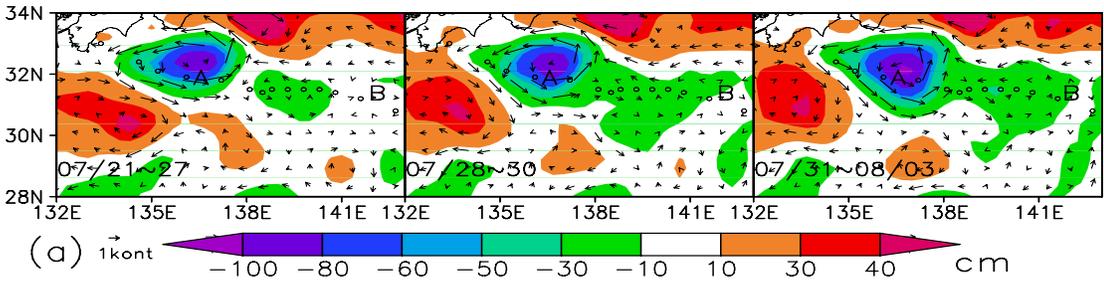

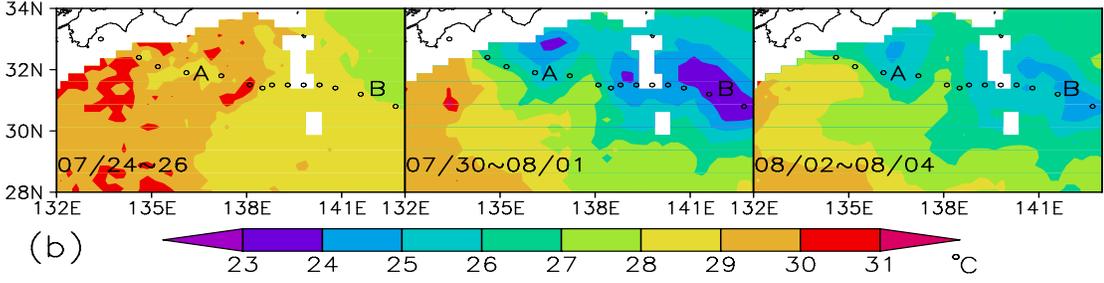

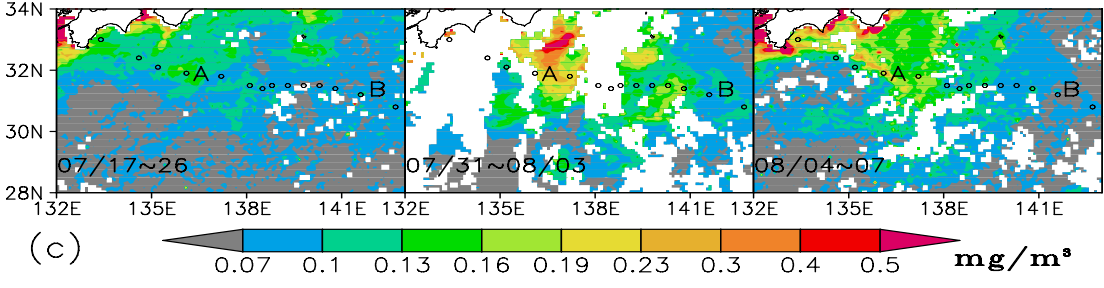

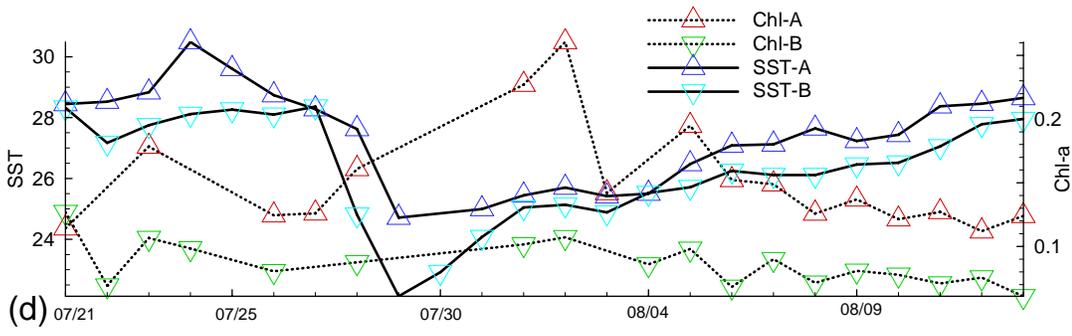

Figure 2

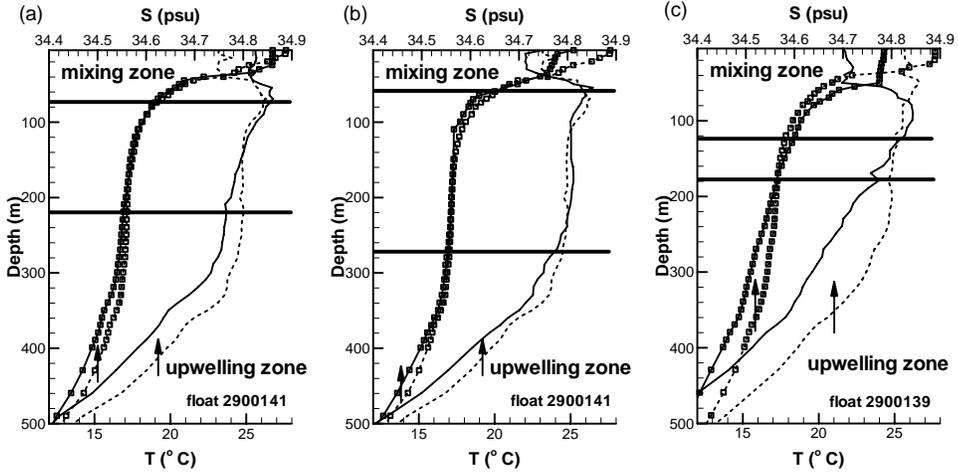

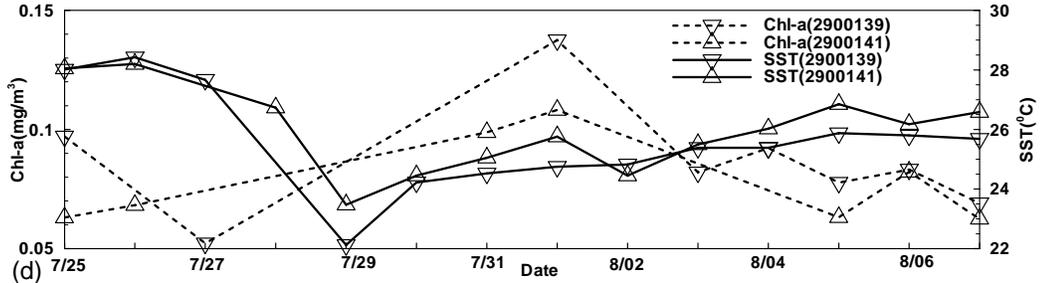

Figure 3